\documentclass{JHEP3}
\usepackage{amsmath}
\usepackage[mathscr]{eucal}

\title{Notes on SUSY Gauge Theories on Three-Sphere}

\author{
Naofumi Hama$^a$, Kazuo Hosomichi$^a$ and Sungjay Lee$^b$\\
$^a$Yukawa Institute for Theoretical Physics, Kyoto University,\\
~  Kitashirakawa-Oiwakecho, Sakyo-ku, Kyoto 606-8502, Japan
\vskip2mm
$^b$DAMTP, Centre for Mathematical Sciences, Cambridge University,\\
~  Wilberforce Road, Cambridge, CB3 0WA, UK
}

\abstract{
We extend the formula for partition functions of ${\cal N}=2$
superconformal gauge theories on $S^3$ obtained recently by
Kapustin, Willett and Yaakov, to incorporate matter fields with
arbitrary R-charge assignments.
We use the result to check that the self-mirror property of
${\cal N}=4$ SQED with two electron hypermultiplets is preserved under
a certain mass deformation which breaks the supersymmetry to ${\cal N}=2$.}

\preprint{DAMTP-2010-129 \\ YITP-10-100}


\begin{document}

\section{Introduction}\label{sec:intro}

A major progress in the area of supersymmetric gauge theories has been
made in recent years based on exact computation of path integral on some
deformed or compact manifolds. In four dimensions, it was shown in the
pioneering work \cite{Nek} that the exact ${\cal N}=2$ prepotential
can be extracted from the path integral on Omega-deformed spacetime.
In a similar manner, based on the localization principle, the partition
function and Wilson loop observables of Seiberg-Witten theories on $S^4$
have been computed in \cite{Pe}. These results led to a discovery of a
remarkable relation between 4D gauge theories and 2D Liouville or Toda
CFTs\cite{AGT,Wy}, called AGT relation.

For 3D ${\cal N}=2$ superconformal gauge theories on $S^3$, exact
partition functions and Wilson loop observables have been obtained
in \cite{KWY1}. The techniques developed there have been applied to
further studies of various topics, such as Wilson loops \cite{Suy,RS},
3D dualities \cite{KWY2} and large-$N$ duality of topological string
\cite{MPP}. It has also been applied to the study of the ABJM theory
at strong coupling, in particular its conjectured ${\cal O}(N^{3/2})$
growth of the degrees of freedom\cite{DT,MP,DMP}. Another application
has recently been made to the study of domain walls in 4D ${\cal N}=2$
gauge theories \cite{DGG,HLP} in connection with the AGT relation.

The path integration of fields was performed in \cite{KWY1} for gauge
theories with manifestly superconformally invariant Lagrangian.
In particular, all the matter scalars and fermions are assigned canonical
dimensions $1/2$ and $1$, respectively. Using the same technique,
the partition functions of various ${\cal N}=4$ superconformal
gauge theories was computed in \cite{KWY2} as functions of the relevant
(FI and mass) deformation parameters. One subtle issue there was that
an ${\cal N}=4$ vectormultiplet contains an ${\cal N}=2$ chiral multiplet
with non-canonical dimension. In \cite{KWY2}, the contribution from such
chiral multiplet to the partition function was argued to be trivial,
by pointing out the existence of a SUSY-exact F-term
deformation which lifts all of its component fields. One should be able
to check this by more direct means. Also, it remained unclear whether this
property continues to hold when a mass to this chiral matter is
turned on to break supersymmetry to ${\cal N}=2$.

In this paper we extend the result of \cite{KWY1,KWY2} so that the matter
chiral multiplets with arbitrary R-charge assignment can be incorporated.
After summarizing in Section \ref{sec:three-sphere} our notations for
various geometric quantities in $S^3$, we give a supersymmetry
transformation law of ${\cal N}=2$ vector and chiral multiplets in
Section \ref{sec:SUSY}. There we also construct various supersymmetric
Lagrangians; among them are the super Yang-Mills Lagrangian
for vectormultiplets and kinetic Lagrangian for chiral multiplets.
Similar Lagrangians were studied in the context of 4D ${\cal N}=1$
gauge theories on $S^3\times \mathbb R$ in \cite{Sen,Rom}.
They are both shown to be total superderivatives, so it follows that the
partition function does not depend on the Yang-Mills coupling.
Then, in Section \ref{sec:localization} we compute the one-loop
determinant of general chiral matters on the saddle points parametrized
by the vev of vectormultiplet scalars. The prescription to compute
partition function for general ${\cal N}=2$ gauge theories on $S^3$
is summarized in Section \ref{sec:general}. Finally, in Section
\ref{sec:application} we apply our result to check the self-mirror
property of a certain ${\cal N}=2$ SQED which has recently been studied
in \cite{HLP}.

\section{Three-Sphere}\label{sec:three-sphere}

The three-sphere is parametrized by an element $g$ of the Lie group
$SU(2)$, and two copies of $SU(2)$ symmetry act on $g$ from the left
and the right. We introduce the left-invariant (LI) and right-invariant
(RI) one-forms
$\mu^a=\mu^a_\nu d\xi^\nu$ and $\tilde\mu^a=\tilde\mu^a_\nu d\xi^\nu$,
\begin{equation}
 g^{-1}dg ~=~ i\mu^a\gamma^a,\quad
 dg g^{-1}~=~ i\tilde\mu^a\gamma^a,
\end{equation}
where $\gamma^a$ are Pauli matrices.
These one-forms satisfy
\begin{equation}
 d\mu^a=\epsilon^{abc}\mu^b\mu^c,\quad
 d\tilde\mu^a=-\epsilon^{abc}\tilde\mu^b\tilde\mu^c.
\end{equation}
The left-right invariant round metric with radius $\ell$ is
\begin{equation}
 ds^2 ~=~ \tfrac12\ell^2\text{tr}(dgdg^{-1})
 ~=~ \ell^2\mu^a\mu^a~=~\ell^2\tilde\mu^a\tilde\mu^a.
\end{equation}
We define the vielbein in the ``LI frame'' as
$e^a = e^a_\mu d\xi^\mu = \ell\mu^a$.
The spin connection in this frame is
$\omega^{ab}= \varepsilon^{abc}\mu^c$ and satisfies
$de^a+\omega^{ab}e^b=0$.
If we define the vielbein from $\tilde\mu^a$ (``RI frame''),
the spin connection is $\tilde\omega^{ab}=-\varepsilon^{abc}\tilde\mu^c$.

\paragraph{Killing spinors.}
Killing spinor $\epsilon$ satisfies the following equation
\begin{equation}
 D\epsilon
 ~\equiv~ d\epsilon+\tfrac14\gamma^{ab}\omega^{ab}\epsilon
 ~=~ e^a\gamma^a\tilde\epsilon,
\label{KSeq}
\end{equation}
for a certain $\tilde\epsilon$. Here we used the notation
$\gamma^{ab}\equiv\frac12[\gamma^a,\gamma^b]=i\varepsilon^{abc}\gamma^c$.
There are two types of Killing spinors. The first one is constant in the
LI frame,
\begin{equation}
 \epsilon=\epsilon_0~(\text{constant}), \quad
 \tilde\epsilon=+\tfrac i{2\ell}\epsilon.
\end{equation}
The second one reads
\begin{equation}
 \epsilon=g^{-1}\epsilon_0, \quad
 \tilde\epsilon=-\tfrac i{2\ell}\epsilon,
\end{equation}
and is constant in the RI frame.

\paragraph{Killing vectors.}

Let us next introduce the vector fields
$\mathscr{L}^a=\mathscr{L}^{a\mu}\frac\partial{\partial\xi^\mu}$ and
$\mathscr{R}^a=\mathscr{R}^{a\mu}\frac\partial{\partial\xi^\mu}$
which generate the left and the right actions of $SU(2)$.
They can be determined from
\begin{equation}
 \mathscr{L}^ag= i\gamma^ag,\quad
 \mathscr{R}^ag= ig\gamma^a.
\end{equation}
The vector fields $\tfrac i2\mathscr{L}^a$ and $-\tfrac i2\mathscr{R}^a$
satisfy the standard commutation relations of $SU(2)$ Lie algebra.
It is also easy to find
$\mathscr{R}^{a\nu}\mu_\nu^b=\mathscr{L}^{a\nu}\tilde\mu_\nu^b=\delta^{ab}$,
in other words $\mathscr{R}^{a\nu}$ and $\mathscr{L}^{a\nu}$ are proportional
to the inverse-vielbeins in LI or RI frames.
The action of these Killing vector fields on the LI and RI one-forms
is given by
\begin{equation}
 \mathscr{L}^a\tilde\mu^b=2\varepsilon^{abc}\tilde\mu^c,\quad
 \mathscr{R}^a\mu^b=-2\varepsilon^{abc}\mu^c,\quad
 \mathscr{L}^a\mu^b=\mathscr{R}^a\tilde\mu^b=0.
\end{equation}
It therefore follows that $\mu^1\mu^2\mu^3= d^3\xi\text{det}(\mu^a_\nu)$
can be used to define the invariant volume form.

\section{SUSY Theories on Three-Sphere}\label{sec:SUSY}

Here we review the construction of Euclidean 3D ${\cal N}=2$
superconformal gauge theories on manifolds with Killing spinors \cite{KWY1},
and extend it to non-conformal theories.
We begin by summarizing our conventions for bilinear
products of spinors.
\begin{equation}
 \bar\epsilon\lambda~=~ \bar\epsilon^\alpha C_{\alpha\beta}\lambda^\beta,\quad
 \bar\epsilon\gamma^\mu\lambda~=~
 \bar\epsilon^\alpha(C\gamma^\mu)_{\alpha\beta}\lambda^\beta,\quad
 \text{etc}.
\end{equation}
Here $C$ is the charge conjugation matrix.
Noticing that $C$ is antisymmetric and $C\gamma_\mu$ are symmetric,
one finds
\begin{equation}
 \bar\epsilon\lambda=\lambda\bar\epsilon,\quad
 \bar\epsilon\gamma^\mu\lambda=-\lambda\gamma^\mu\bar\epsilon
\end{equation}
for all spinors $\bar\epsilon,\lambda$ which we assume to be Grassmann
odd.

\paragraph{Vectormultiplets.}

The vectormultiplet fields obey the following transformation laws,
\begin{eqnarray}
 \delta A_\mu &=&
 -\tfrac i2(\bar\epsilon\gamma_\mu\lambda-\bar\lambda\gamma_\mu\epsilon),
 \nonumber \\
 \delta\sigma &=&
 \tfrac12(\bar\epsilon\lambda-\bar\lambda\epsilon),
 \nonumber \\
 \delta\lambda &=&
 \tfrac12\gamma^{\mu\nu}\epsilon F_{\mu\nu}-D\epsilon
      +i\gamma^\mu\epsilon D_\mu\sigma
 +\tfrac{2i}3\sigma\gamma^\mu D_\mu\epsilon,
 \nonumber \\
 \delta\bar\lambda &=&
 \tfrac12\gamma^{\mu\nu}\bar\epsilon F_{\mu\nu}+D\bar\epsilon
                 -i\gamma^\mu\bar\epsilon D_\mu\sigma
 -\tfrac{2i}3\sigma\gamma^\mu D_\mu\bar\epsilon,
 \nonumber \\
 \delta D &=&
 -\tfrac i2\bar\epsilon\gamma^\mu D_\mu\lambda
 -\tfrac i2D_\mu\bar\lambda\gamma^\mu\epsilon
 +\tfrac i2[\bar\epsilon\lambda,\sigma]
 +\tfrac i2[\bar\lambda\epsilon,\sigma]
 -\tfrac i6(D_\mu\bar\epsilon\gamma^\mu\lambda
         +\bar\lambda\gamma^\mu D_\mu\epsilon).
\label{trvec}
\end{eqnarray}
Here and throughout this paper, $D_\mu$ denotes the gauge, local Lorentz
and general covariant derivative, and $\gamma^\mu$ is the Dirac matrix
with curved index which satisfies
\begin{equation}
 \{\gamma_\mu,\gamma_\nu\}=2g_{\mu\nu},\quad
 \gamma^{\mu\nu}=i\varepsilon^{\mu\nu\rho}\gamma_\rho/\sqrt{g}.\quad
 (\varepsilon^{123}=1)
\end{equation}
Note that $D_\mu$ commutes with the vielbein $e_\mu^a$ and the
Dirac matrices $\gamma^a$ or $\gamma^\mu$.
The spinors $\epsilon,\bar\epsilon$ are assumed to satisfy
Killing spinor equation.
Denoting $\delta$ as the sum of unbarred and barred parts,
$\delta=\delta_\epsilon+\delta_{\bar\epsilon}$, one can show that
two unbarred or two barred supersymmetries commute. Also, on all the
fields except $D$ the commutator $[\delta_\epsilon,\delta_{\bar\epsilon}]$
becomes a sum of translation, gauge transformation, Lorentz rotation,
dilation and R-rotation.
\begin{eqnarray}
 ~[\delta_\epsilon,\delta_{\bar\epsilon}]A_\mu &=&
 \xi^\nu\partial_\nu A_\mu + \partial_\mu\xi^\nu A_\nu
 +D_\mu\Lambda,
 \nonumber \\
 ~[\delta_\epsilon,\delta_{\bar\epsilon}]\sigma &=&
 \xi^\mu\partial_\mu\sigma+i[\Lambda,\sigma]+\rho\sigma,
 \nonumber \\
 ~[\delta_\epsilon,\delta_{\bar\epsilon}]\lambda &=&
 \xi^\mu \partial_\mu\lambda+\tfrac14\Theta_{\mu\nu}\gamma^{\mu\nu}\lambda
 +i[\Lambda,\lambda]+\tfrac32\rho\lambda
 +\alpha\lambda,
 \nonumber \\
 ~[\delta_\epsilon,\delta_{\bar\epsilon}]\bar\lambda &=&
 \xi^\mu \partial_\mu\bar\lambda
 +\tfrac14\Theta_{\mu\nu}\gamma^{\mu\nu}\bar\lambda
 +i[\Lambda,\bar\lambda]+\tfrac32\rho\bar\lambda
 -\alpha\bar\lambda,
 \nonumber \\
 ~[\delta_\epsilon,\delta_{\bar\epsilon}]D &=&
 \xi^\mu\partial_\mu D+i[\Lambda,D]+2\rho D \nonumber \\ && \hskip10mm
 +\tfrac13\sigma(\bar\epsilon\gamma^\mu\gamma^\nu D_\mu D_\nu\epsilon
                -\epsilon\gamma^\mu\gamma^\nu D_\mu D_\nu\bar\epsilon),
\end{eqnarray}
where
\begin{eqnarray}
 \xi^\mu &=& i\bar\epsilon\gamma^\mu\epsilon,
 \nonumber \\
 \Theta^{\mu\nu} &=& D^{[\mu}\xi^{\nu]}+\xi^\lambda\omega_\lambda^{\mu\nu},
 \nonumber \\
 \Lambda &=& -iA_\mu\bar\epsilon\gamma^\mu\epsilon
           +\sigma\bar\epsilon\epsilon,
 \nonumber \\
 \rho &=& \tfrac i3
 (\bar\epsilon\gamma^\mu D_\mu\epsilon
 +D_\mu\bar\epsilon\gamma^\mu\epsilon),
 \nonumber \\
 \alpha &=& \tfrac
  i3(D_\mu\bar\epsilon\gamma^\mu\epsilon-\bar\epsilon\gamma^\mu D_\mu\epsilon).
\label{sypar}
\end{eqnarray}
In order for the supersymmetry algebra to close, the last term in the
right hand side of $[\delta_\epsilon,\delta_{\bar\epsilon}]D$ needs to
vanish. The Killing spinors therefore have to satisfy, in addition to
(\ref{KSeq}), the following condition
\begin{equation}
 \gamma^\mu\gamma^\nu D_\mu D_\nu\epsilon = h\epsilon,
\label{Kil2}
\end{equation}
with some scalar function $h$.
The barred spinor $\bar\epsilon$ also has to satisfy the same
equation with the same $h$.
By combining this with Killing spinor equation (\ref{KSeq}), one obtains
\begin{equation}
 D_\mu\epsilon=\gamma_\mu\tilde\epsilon,\quad
 3\gamma^\mu D_\mu\tilde\epsilon=h\epsilon.
\end{equation}
By inserting this into $\gamma^{\mu\nu}D_\mu D_\nu\epsilon=-\tfrac14R\epsilon$
one finds $h=-\frac{3R}8$, where $R$ is the scalar curvature of the
3D manifold. For $S^3$ of radius $\ell$ one has $R=\frac6{\ell^2}$
and therefore
\begin{equation}
h=-\tfrac{9}{4\ell^2}.
\label{Kil3}
\end{equation}
Note that all the Killing spinors on $S^3$ satisfy
$D_\mu\epsilon=\pm\frac i{2\ell}\gamma_\mu\epsilon$, so they
automatically satisfy the additional condition (\ref{Kil2}).
So the additional condition does not reduce the number of supersymmetry
on $S^3$.

The parameters $\rho,\alpha$ are associated to dilation and R-rotation,
respectively. The above result shows that the fields $(A_\mu,\sigma,\lambda,D)$
have dimensions $(1,1,3/2,2)$, and $(\lambda,\bar\lambda)$ are assigned
the R-charge $(1,-1)$.

\paragraph{Matter multiplets.}

The fields in a chiral multiplet coupled to a gauge symmetry
transform as follows,
\begin{eqnarray}
 \delta\phi &=& \bar\epsilon\psi, \nonumber \\
 \delta\bar\phi &=& \epsilon\bar\psi, \nonumber \\
 \delta\psi &=& i\gamma^\mu\epsilon D_\mu\phi +i\epsilon\sigma\phi
 +\tfrac{i}3\gamma^\mu D_\mu\epsilon\phi+\bar\epsilon F,
 \nonumber \\
 \delta\bar\psi &=& i\gamma^\mu\bar\epsilon D_\mu\bar\phi
 +i\bar\phi\sigma\bar\epsilon+\tfrac{i}3\bar\phi\gamma^\mu D_\mu\bar\epsilon
 +\bar F\epsilon,
 \nonumber \\
 \delta F &=&
 \epsilon(i\gamma^\mu D_\mu\psi-i\sigma\psi-i\lambda\phi),
 \nonumber \\
 \delta\bar F &=&
 \bar\epsilon(i\gamma^\mu D_\mu\bar\psi-i\bar\psi\sigma+i\bar\phi\bar\lambda).
\end{eqnarray}
Here we assumed the fields $\phi,\psi,F$ ($\bar\phi,\bar\psi,\bar F$) to be
column vectors (resp. row vectors) on which the vectormultiplet fields
act as matrices from the left (right).
The supersymmetry algebra closes off-shell. The commutator
$[\delta_\epsilon,\delta_{\bar\epsilon}]$ on matter fields reads
\begin{eqnarray}
~[\delta_\epsilon,\delta_{\bar\epsilon}]\phi &=&
 \xi^\mu\partial_\mu\phi+i\Lambda\phi+\tfrac\rho2\phi-\tfrac\alpha2\phi,
 \nonumber \\
~[\delta_\epsilon,\delta_{\bar\epsilon}]\bar\phi &=&
 \xi^\mu\partial_\mu\bar\phi-i\bar\phi\Lambda+\tfrac\rho2\bar\phi
 +\tfrac\alpha2\bar\phi,
 \nonumber \\
~[\delta_\epsilon,\delta_{\bar\epsilon}]\psi &=&
 \xi^\mu \partial_\mu\psi
 +\tfrac14\Theta_{\mu\nu}\gamma^{\mu\nu}\psi
 +i\Lambda\psi+\rho\psi+\tfrac\alpha2\psi,
 \nonumber \\
~[\delta_\epsilon,\delta_{\bar\epsilon}]\bar\psi &=&
 \xi^\mu \partial_\mu\bar\psi
 +\tfrac14\Theta_{\mu\nu}\gamma^{\mu\nu}\bar\psi
 -i\bar\psi\Lambda+\rho\bar\psi
 -\tfrac\alpha2\bar\psi,
 \nonumber \\
~[\delta_\epsilon,\delta_{\bar\epsilon}]F &=&
 \xi^\mu\partial_\mu F+i\Lambda F+\tfrac{3\rho}2F+\tfrac{3\alpha}2F,
 \nonumber \\
~[\delta_\epsilon,\delta_{\bar\epsilon}]\bar F &=&
 \xi^\mu\partial_\mu\bar F-i\bar F\Lambda+\tfrac{3\rho}2\bar F
 -\tfrac{3\alpha}2\bar F.
\end{eqnarray}
This shows that the fields $(\phi,\psi,F)$ are assigned the canonical
dimensions $(1/2,1,3/2)$.
Two unbarred or two barred supersymmetries can be easily shown to
commute, except on the auxiliary fields. On $F$ one finds
\begin{equation}
 [\delta_\epsilon,\delta_{\epsilon'}]F ~=~
 \tfrac13\phi\cdot
 (\epsilon \gamma^\mu\gamma^\nu D_\mu D_\nu\epsilon'
 -\epsilon'\gamma^\mu\gamma^\nu D_\mu D_\nu\epsilon),
\end{equation}
which vanishes if $\epsilon,\epsilon'$ satisfy the constraint (\ref{Kil2}).
Similarly, the commutator of two barred supersymmetries vanish on
$\bar F$ only if the two barred Killing spinors satisfy the same constraint.

For matter multiplets with non-canonical assignments of dimensions,
we put the supersymmetry transformation rule as follows,
\begin{eqnarray}
 \delta\phi &=& \bar\epsilon\psi, \nonumber \\
 \delta\bar\phi &=& \epsilon\bar\psi, \nonumber \\
 \delta\psi &=& i\gamma^\mu\epsilon D_\mu\phi +i\epsilon\sigma\phi
 +\tfrac{2qi}3\gamma^\mu D_\mu\epsilon\phi+\bar\epsilon F,
 \nonumber \\
 \delta\bar\psi &=& i\gamma^\mu\bar\epsilon D_\mu\bar\phi
 +i\bar\phi\sigma\bar\epsilon+\tfrac{2qi}3\bar\phi\gamma^\mu D_\mu\bar\epsilon
 +\bar F\epsilon,
 \nonumber \\
 \delta F &=&
 \epsilon(i\gamma^\mu D_\mu\psi-i\sigma\psi-i\lambda\phi)
 +\tfrac i3(2q-1)D_\mu\epsilon\gamma^\mu\psi,
 \nonumber \\
 \delta\bar F &=&
 \bar\epsilon(i\gamma^\mu D_\mu\bar\psi-i\bar\psi\sigma+i\bar\phi\bar\lambda)
 +\tfrac i3(2q-1)D_\mu\bar\epsilon\gamma^\mu\bar\psi.
\label{dncm}
\end{eqnarray}
The supersymmetry algebra then becomes
\begin{eqnarray}
~[\delta_\epsilon,\delta_{\bar\epsilon}]\phi &=&
 \xi^\mu\partial_\mu\phi+i\Lambda\phi+q\rho\phi-q\alpha\phi,
 \nonumber \\
~[\delta_\epsilon,\delta_{\bar\epsilon}]\bar\phi &=&
 \xi^\mu\partial_\mu\bar\phi-i\bar\phi\Lambda+q\rho\bar\phi
 +q\alpha\bar\phi,
 \nonumber \\
~[\delta_\epsilon,\delta_{\bar\epsilon}]\psi &=&
 \xi^\mu \partial_\mu\psi
 +\tfrac14\Theta_{\mu\nu}\gamma^{\mu\nu}\psi
 +i\Lambda\psi+(q+\tfrac12)\rho\psi+(1-q)\alpha\psi,
 \nonumber \\
~[\delta_\epsilon,\delta_{\bar\epsilon}]\bar\psi &=&
 \xi^\mu \partial_\mu\bar\psi
 +\tfrac14\Theta_{\mu\nu}\gamma^{\mu\nu}\bar\psi
 -i\bar\psi\Lambda+(q+\tfrac12)\rho\bar\psi
 +(q-1)\alpha\bar\psi,
 \nonumber \\
~[\delta_\epsilon,\delta_{\bar\epsilon}]F &=&
 \xi^\mu\partial_\mu F+i\Lambda F+(q+1)\rho F+(2-q)\alpha F,
 \nonumber \\
~[\delta_\epsilon,\delta_{\bar\epsilon}]\bar F &=&
 \xi^\mu\partial_\mu\bar F-i\bar F\Lambda+(q+1)\rho\bar F
 +(q-2)\alpha\bar F.
\end{eqnarray}
The lowest components are now assigned the dimension $q$ and R-charge
$\mp q$.
The supersymmetry algebra closes off-shell when the Killing spinors
$\epsilon,\bar\epsilon$ satisfy (\ref{Kil2}), (\ref{Kil3}).

\paragraph{Supersymmetric Lagrangians.}

The Chern-Simons Lagrangian for ${\cal N}=2$ vectormultiplet is invariant
under supersymmetry.
\begin{equation}
 {\cal L}_{\rm CS} ~=~ {\rm Tr}\left[
   \tfrac1{\sqrt g}\varepsilon^{\mu\nu\lambda}
  (A_\mu\partial_\nu A_\lambda-\tfrac{2i}3A_\mu A_\nu A_\lambda)
   -\bar\lambda\lambda+2D\sigma  \right].
\label{LCS}
\end{equation}
Given a gauge-invariant chiral multiplet of R-charge $q=2$ usually called
{\it superpotential}, its F-term is invariant under supersymmetry
up to total derivatives.
\begin{equation}
 \delta F ~=~ iD_\mu(\epsilon\gamma^\mu\psi),\quad
 \delta\bar F ~=~ iD_\mu(\bar\epsilon\gamma^\mu\bar\psi).
\end{equation}
These terms are invariant under $\delta$ for any Killing spinors
$\epsilon,\bar\epsilon$.
In addition, chiral matter multiplets with canonical dimensions have
the kinetic Lagrangian,
\begin{eqnarray}
 {\cal L} &=& D_\mu\bar\phi D^\mu\phi -i\bar\psi\gamma^\mu D_\mu\psi
 +\tfrac3{4\ell^2}\bar\phi\phi+i\bar\psi\sigma\psi
 \nonumber \\ &&
 +i\bar\psi\lambda\phi-i\bar\phi\bar\lambda\psi
 +i\bar\phi D\phi+\bar\phi\sigma^2\phi+\bar FF,
\label{Lcm}
\end{eqnarray}
which is invariant under supersymmetry if the Killing spinors
$\epsilon,\bar\epsilon$ satisfy (\ref{Kil2}), (\ref{Kil3}).
If the Lagrangian is made of the above three types of terms, the theory
is superconformal at the classical level.

There are Lagrangians which are not superconformal but are
still invariant under some supersymmetry. In the following we look for
the quantities which are invariant if the parameters
$\epsilon,\bar\epsilon$ satisfy
\begin{equation}
 D_\mu\epsilon=\tfrac{i}{2\ell}\gamma_\mu\epsilon,\quad
 D_\mu\bar\epsilon=\tfrac{i}{2\ell}\gamma_\mu\bar\epsilon.
\label{nscd}
\end{equation}
Under this additional condition, the commutator
$[\delta_\epsilon,\delta_{\bar\epsilon}]$ does not give rise to
dilation since $\rho$ of (\ref{sypar}) vanishes.
The Killing vector $\bar\epsilon\gamma^a\epsilon$ is constant in
the LI frame, so the commutator of supersymmetry is a linear sum
of $\mathscr R^a$ and local Lorentz, gauge, R-transformations.
This restricted supersymmetry is therefore regarded as an analogue of
Poincar\'e supersymmetry in flat space.

As an example, let us first look for a kinetic Lagrangian for
matter fields with non-canonical dimensions. We take (\ref{Lcm}) as
the trial Lagrangian for non-canonical matters. Its variation under
the supersymmetry (\ref{dncm}) is given by
\begin{eqnarray}
 \delta{\cal L} &=&
 -\tfrac i3(2q-1)\bar\phi D_\mu\bar\epsilon\gamma^\mu
 (-i\gamma^\nu D_\nu\psi+i\sigma\psi+i\lambda\phi)
 \nonumber \\&&
 +\tfrac i3(2q-1)(iD_\mu\bar\psi\gamma^\mu+i\bar\psi\sigma-i\bar\phi\bar\lambda)
  \gamma^\nu D_\nu\epsilon\phi
 \nonumber \\&&
 +\tfrac i3(2q-1)(\bar F D_\mu\epsilon\gamma^\mu\psi
 -\bar\psi\gamma^\mu D_\mu\bar\epsilon F).
\end{eqnarray}
Using (\ref{nscd}) one can rewrite this as
$\delta{\cal L}=-\delta{\cal L}_\text{nc}$, where
\begin{equation}
 {\cal L}_\text{nc} ~=~
 \tfrac{i(2q-1)}\ell\bar\phi\sigma\phi
 -\tfrac{(2q-1)}{2\ell}\bar\psi\psi
 -\tfrac{(2q-1)(2q-3)}{4\ell^2}\bar\phi\phi.
\end{equation}
Thus ${\cal L}_\text{mat}={\cal L}+{\cal L}_\text{nc}$ is a
supersymmetric kinetic Lagrangian.
\begin{eqnarray}
 {\cal L}_\text{mat} &=&
  D_\mu\bar\phi D^\mu\phi
 +\bar\phi\sigma^2\phi
 +\tfrac{i(2q-1)}\ell\bar\phi\sigma\phi
 +\tfrac{q(2-q)}{\ell^2}\bar\phi\phi
 +i\bar\phi D\phi
 +\bar FF
 \nonumber \\ &&
 -i\bar\psi\gamma^\mu D_\mu\psi
 +i\bar\psi\sigma\psi
 -\tfrac{(2q-1)}{2\ell}\bar\psi\psi
 +i\bar\psi\lambda\phi
 -i\bar\phi\bar\lambda\psi.
\label{Lncm}
\end{eqnarray}
Another example is the Yang-Mills Lagrangian for vectormultiplet.
We start from the standard ${\cal N}=2$ SYM Lagrangian and improve
its supersymmetry variation by adding terms of order $\ell^{-1}$
and $\ell^{-2}$.
The supersymmetric Lagrangian finally becomes
\begin{eqnarray}
 {\cal L}_\text{YM} &=& \text{Tr}\Big(
  \tfrac14F_{\mu\nu}F^{\mu\nu}+\tfrac12D_\mu\sigma D^\mu\sigma
 +\tfrac12(D+\tfrac\sigma\ell)^2
 \nonumber \\ && ~~~
 +\tfrac i2\bar\lambda\gamma^\mu D_\mu\lambda
 +\tfrac i2\bar\lambda[\sigma,\lambda]
 -\tfrac1{4\ell}\bar\lambda\lambda
 \Big).
\end{eqnarray}
Finally, there is an analogue of FI D-term for abelian vectormultiplet.
\begin{equation}
 {\cal L}_\text{FI}\equiv D-\tfrac\sigma\ell,\quad
 \delta {\cal L}_\text{FI} =
 -\tfrac i2D_\mu(\bar\epsilon\gamma^\mu\lambda
                       +\bar\lambda\gamma^\mu\epsilon).
\label{LFI}
\end{equation}

Note that ${\cal L}_\text{mat}$ and ${\cal L}_\text{YM}$ can be
expressed as total-superderivatives,
\begin{eqnarray}
 \bar\epsilon\epsilon\cdot{\cal L}_\text{mat} &=&
 \delta_{\bar\epsilon}\delta_\epsilon\Big(
  \bar\psi\psi-2i\bar\phi\sigma\phi+\tfrac{2(q-1)}{\ell}\bar\phi\phi
 \Big),
 \nonumber \\
 \bar\epsilon\epsilon\cdot{\cal L}_\text{YM} &=&
 \delta_{\bar\epsilon}\delta_\epsilon\text{Tr}\Big(
  \tfrac12\bar\lambda\lambda-2D\sigma \Big),
\end{eqnarray}
but ${\cal L}_\text{FI}$ cannot.

\paragraph{More extended supersymmetry.}

By combining a vectormultiplet with an adjoint chiral multiplet we
expect to get a gauge theory with more extended supersymmetry.
In the following we will write the Lagrangians for this extended
multiplet using a new set of fields,
\begin{equation}
 \hat\phi=\sqrt2\phi,\quad
 \hat\psi=-i\sqrt2\psi,\quad
 \hat{\bar\psi}=-i\sqrt2\bar\psi,\quad
 \hat F=\sqrt2F,\quad
 \hat D=D+i[\phi,\bar\phi].
\end{equation}

To get the Lagrangian with ${\cal N}=4$ extended supersymmetry,
it turns out one has to take a linear combination of YM, matter
and CS terms.
The total Lagrangian becomes (hereafter the hats for the new set of
fields are omitted),
\begin{eqnarray}
\lefteqn{
 {\cal L}_\text{YM}+{\cal L}_\text{mat}-\tfrac1{2\ell}{\cal L}_\text{CS}
} \nonumber \\
 &=&
\text{Tr}\Big(
 \tfrac14F_{\mu\nu}F^{\mu\nu}-\tfrac1{2\ell\sqrt g}\varepsilon^{\mu\nu\lambda}
 (A_\mu\partial_\nu A_\lambda-\tfrac{2i}3A_\mu A_\nu A_\lambda)
 +\tfrac12D^2+\tfrac12\bar FF
 \nonumber \\ &&\hskip4mm
 +\tfrac12D_\mu\sigma D^\mu\sigma+\tfrac12 D_\mu\bar\phi D^\mu\phi
 +\tfrac1{2\ell^2}(\sigma^2+\bar\phi\phi)
 -\tfrac12[\sigma,\phi][\sigma,\bar\phi]+\tfrac18[\phi,\bar\phi]^2
 +\tfrac i{2\ell}\sigma[\phi,\bar\phi]
 \nonumber \\&&\hskip4mm
 +\tfrac i2\bar\lambda\gamma^\mu D_\mu\lambda
 +\tfrac i2\bar\psi\gamma^\mu D_\mu\psi
 +\tfrac1{4\ell}(\bar\lambda\lambda+\bar\psi\psi)
 \nonumber \\&&\hskip4mm
 +\tfrac i2\bar\lambda[\sigma,\lambda]-\tfrac i2\bar\psi[\sigma,\psi]
 +\tfrac12\bar\psi[\phi,\lambda]-\tfrac12\bar\lambda[\bar\phi,\psi]
\Big).
\end{eqnarray}
This Lagrangian has an $SO(4)\simeq SU(2)\times SU(2)$ enlarged
R-symmetry and therefore ${\cal N}=4$ supersymmetry.
The two $SU(2)$'s act on the triplet of scalars $(\sigma,\phi,\bar\phi)$
and the auxiliary fields $(D,F,\bar F)$ respectively.

By combining the complex mass term (F-term) for the adjoint chiral field
\begin{equation}
 {\cal L}_F+{\cal L}_{\bar F}~=~
 \text{Tr}\big(F\phi+\tfrac12\psi\psi
              +\bar F\bar\phi+\tfrac12\bar\psi\bar\psi\big)
\end{equation}
with the CS action for the vectormultiplet one obtains an action
\begin{eqnarray}
 {\cal L}_\text{CS}+{\cal L}_F+{\cal L}_{\bar F} &=&
 \text{Tr}\Big(
 \tfrac1{\sqrt g}\varepsilon^{\mu\nu\lambda}
 (A_\mu\partial_\nu A_\lambda-\tfrac{2i}3A_\mu A_\nu A_\lambda)
 -\bar\lambda\lambda+\tfrac12\psi\psi+\tfrac12\bar\psi\bar\psi
 \nonumber \\ && \hskip4mm
 +2D\sigma+F\phi+\bar F\bar\phi-i[\phi,\bar\phi]\sigma
 \Big),
\end{eqnarray}
which has an $SO(3)$ extended R-symmetry which rotates the scalars,
auxiliary fields and three of the four Majorana fermions simultaneously,
and therefore ${\cal N}=3$ supersymmetry.

The above observations are reminiscent of the well-known fact that,
on 3D flat spacetime, gauge theories with YM and CS terms can have
at most ${\cal N}=3$ supersymmetry.

\section{Localization}\label{sec:localization}

Here we discuss the computation of partition function of the
supersymmetric gauge theories on $S^3$ based on the
localization principle. As has been explained in \cite{KWY1,KWY2},
the path integral localizes onto the saddle points characterized by
\begin{equation}
 A_\mu = \phi= 0,\quad
 \sigma=-\ell D= \text{constant}.
\label{saddle}
\end{equation}
So the calculation of partition function amounts to evaluating the
one-loop determinant at each saddle point and then integrating over
the space of saddle points parametrized by $\sigma$.

The one-loop determinant of vectormultiplets was worked out thoroughly
in \cite{KWY1}, so we focus on the chiral matter multiplets with
arbitrary R-charge assignments.
We will focus on the determinant of a single chiral multiplet which has
a unit charge under a $U(1)$ gauge symmetry, as the generalization to
arbitrary gauge groups and representations is straightforward.

The matter kinetic Lagrangian (\ref{Lncm}) of the previous section was
shown to be a total superderivative, so that we can use it as the regulator
Lagrangian. This choice of regulator is slightly different from the one
in \cite{KWY1}, and it simplifies the computation of one-loop
determinant a lot since the $SU(2)\times SU(2)$ isometry of $S^3$
remains unbroken.

The matter kinetic term on the saddle points is
${\cal L}_\phi+{\cal L}_\psi$, with
\begin{eqnarray}
 {\cal L}_\phi &=&
 g^{\mu\nu}\partial_\mu\bar\phi\partial_\nu\phi +\bar\phi\sigma^2\phi
 +\tfrac{2i(q-1)}\ell\bar\phi\sigma\phi
 +\tfrac{q(2-q)}{\ell^2}\bar\phi\phi,
 \nonumber \\
 {\cal L}_\psi &=&
 -i\bar\psi\gamma^\mu\partial_\mu\psi
 +i\bar\psi\sigma\psi
 -\tfrac{q-2}{\ell}\bar\psi\psi.
\end{eqnarray}
For round $S^3$ of radius $\ell$ we substitute
$g^{\mu\nu}=\ell^{-2}\mathscr R^{a\mu}\mathscr R^{a\nu},~
 e^{a\mu}=\ell^{-1}\mathscr R^{a\mu}$
and get
\begin{eqnarray}
 {\cal L}_\phi &=&
 \ell^{-2}\left\{
  \mathscr R^a\bar\phi\cdot\mathscr R^a\phi
 -\bar\phi(q-i\ell\sigma)(q-2-i\ell\sigma)\phi
 \right\},
 \nonumber \\
 {\cal L}_\psi &=&
 \ell^{-1}\bar\psi\left\{
 -i\gamma^a\mathscr R^a+i\ell\sigma+2-q
 \right\}\psi.
\end{eqnarray}
One can rewrite them in terms of orbital and spin
angular momentum operators $J^a\equiv\frac1{2i}\mathscr R^a$
and $S^a\equiv \frac12\gamma^a$ satisfying standard $SU(2)$
commutation relations.
Then, all we need to do is to work out the spectrum of the Laplace
and Dirac operators $\Delta_\phi$ and $\Delta_\psi$,
\begin{eqnarray}
 \Delta_\phi &=&
 \frac1{\ell^2}\left\{
  4J^aJ^a -(q-i\ell\sigma)(q-2-i\ell\sigma) \right\},
 \nonumber \\
 \Delta_\psi &=& \frac1\ell\left\{
  4J^aS^a+i\ell\sigma+2-q \right\}.
\end{eqnarray}
These operators are diagonalized by spherical harmonics on $S^3$.
First, scalar spherical harmonics sit in the spin $(j,j)$
representations of $SU(2)_\mathscr{L}\times SU(2)_\mathscr{R}$,
with $2j\in\mathbb Z_{\ge0}$. One therefore gets the eigenvalues
\begin{eqnarray}
  \Delta_\phi &=& \ell^{-2}\Big(4j(j+1)-(q-i\ell\sigma)(q-2-i\ell\sigma)\Big)
 \nonumber \\ &=&
 \ell^{-2}(2j+2+i\ell\sigma-q)(2j-i\ell\sigma+q)
\end{eqnarray}
with multiplicity $(2j+1)^2$.
Second, spinor spherical harmonics sit in the spin $(j,j,\frac12)$
representation of $SU(2)_\mathscr{L}\times SU(2)_\mathscr{R}\times SU(2)_S$,
which can be reorganized into the direct sum
$(j,j+\frac12)\oplus(j,j-\frac12)$ of the subgroup
$SU(2)_\mathscr{L}\times SU(2)_{\mathscr{R}+S}$.
The Dirac operator can be easily shown to take values
\begin{eqnarray}
 \Delta_\psi &=&
 \ell^{-1}\Big(
  2(j\pm\tfrac12)(j\pm\tfrac12+1)-2j(j+1)-\tfrac32+i\ell\sigma+2-q
 \Big)
 \nonumber \\ &=&
 \ell^{-1}(2j+2+i\ell\sigma-q),\quad
 \ell^{-1}(-2j+i\ell\sigma-q)
\end{eqnarray}
on these representations, and the multiplicities are
$(2j+2)(2j+1)$ and $2j(2j+1)$ respectively.
Denoting $n=2j+1$, one can express the one-loop determinant
as an infinite product,
\begin{equation}
 \frac{\text{det}\Delta_\psi}{\text{det}\Delta_\phi}~=~
 \prod_{n>0}\Big(\frac{n+1-q+i\ell\sigma}{n-1+q-i\ell\sigma}\Big)^n
 ~=~ s_{b=1}(i-iq-\ell\sigma).
\label{matdet}
\end{equation}
This is our main result.
Here $s_b(x)$ is the double sine function which has poles at
$x=i(m+\frac12)b+i(n+\frac12)b^{-1}~(m,n\in\mathbb Z_{\ge0})$
and satisfies $s_b(x)=s_{1/b}(x)=s_b(-x)^{-1}$.
It also satisfies the equality
\begin{equation}
 s_b(\tfrac{ib}2-\sigma)s_b(\tfrac{ib}2+\sigma)=\frac1{2\cosh\pi b\sigma}.
\end{equation}
For more detailed explanation on this function, we refer to \cite{KLS,BT}.

\section{Integral Formula for Partition Function}\label{sec:general}

Combining the result of the previous section with those for
vectormultiplet given in \cite{KWY1}, one can write down an integral formula
for partition functions of general 3D ${\cal N}=2$ gauge theories.

Since the path integral generally localizes onto saddle points
characterized by (\ref{saddle}), the formula involves an integral over
the Lie algebra of gauge group $G$ corresponding to the constant mode of
$\sigma$. Using gauge symmetry, one can reduce
the integration domain further to its Cartan part at the cost of having an
extra Vandermonde determinant factor in the integrand. We introduce
a dimensionless quantity $\hat\sigma\equiv\ell\sigma$, and write
it as a linear combination of Cartan generators $H_i$,
\begin{equation}
 \hat\sigma= \sum_{i=1}^r\hat\sigma_iH_i,
\end{equation}
where $r$ is the rank of the gauge group.

For non-abelian gauge groups, there is a nontrivial integrand arising
from the Vandermonde and one-loop determinants. The contribution of
vectormultiplets is given by\cite{KWY1}
\begin{equation}
 Z_\text{vec}
 ~=~ \frac1{|{\cal W}|}\int d^r\hat\sigma \prod_{\alpha\in\Delta_+}
 \left(2\sinh(\pi\alpha_i\hat\sigma_i)\right)^2.
\end{equation}
Here $\alpha$ labels the positive roots, the corresponding generator
$E_\alpha$ satisfies $[H_i,E_\alpha]=\alpha_iE_\alpha$ and $|\cal W|$
denotes the order of the Weyl group.

Matter chiral multiplets contribute a one-loop determinant which is a
generalization of (\ref{matdet}). Assume they have R-charge $q$ and
belong to the representation $R$ of the gauge group. Then for each
weight vector $\rho$ of $R$, there is a matter chiral multiplet labelled by
$\rho$ carrying the $H_i$-charge $\rho_i$, and its conjugate anti-chiral
multiplet with the $H_i$-charge $-\rho_i$. Collecting their one-loop
determinants we obtain
\begin{equation}
 \prod_{\rho\in R}s_{b=1}(i-iq-\rho_i\hat\sigma_i)
\end{equation}
from a chiral multiplet belonging to $R$. As a special case, when
$q=\frac12$ and $R={\rm r}\oplus\bar{\rm r}$ the matter one-loop
determinant becomes\cite{KWY1}
\begin{equation}
 \prod_{\rho\in {\rm r}}
 s_{b=1}(\tfrac i2-\rho_i\hat\sigma_i)\cdot
 s_{b=1}(\tfrac i2+\rho_i\hat\sigma_i)
 ~=~
 \prod_{\rho\in {\rm r}}
\frac1{2\cosh\pi\rho_i\hat\sigma_i}\,.
\end{equation}

The Chern-Simons and FI terms have nonzero classical values at the saddle
points. In the standard convention, the Chern-Simons Lagrangian
(\ref{LCS}) appears in the Euclidean action multiplied by $\frac{ik}{4\pi}$,
 where $k$ is the Chern-Simons coupling. Also, the trace in (\ref{LCS})
is that of adjoint representation divided by twice the dual Coxeter
number. Its nonzero classical value shifts the integrand by a factor
\begin{equation}
  \exp\left(\frac{ik}{4\pi}\int d^3\xi\sqrt{g}{\cal L}_{\rm CS}\right)
 ~=~ \exp(-ik\pi\hat\sigma_i\hat\sigma_i).
\end{equation}
The FI term (\ref{LFI}) in our convention appears in the action
multiplied by a factor $\frac{i\zeta}{\pi\ell}$, where $\zeta$ is the FI
coupling. For $U(1)$ gauge theory, the shift of the integrand due to the
FI coupling is therefore
\begin{equation}
 \exp\left(-\frac{i\zeta}{\pi\ell}\int d^3\xi\sqrt{g}{\cal L}_\text{FI}\right)
~=~ \exp(4\pi i\zeta\hat\sigma).
\end{equation}

It is now straightforward to write down the partition function for any
${\cal N}=2$ gauge theories using the building blocks given above. For
example, the partition function for $U(N)$ ${\cal N}=2$ Chern-Simons
theory at level $k$ coupled to $N_f$ fundamental and $\bar N_f$
anti-fundamental chiral matters of R-charge $q$ is,
\begin{eqnarray}
 Z &=& \frac1{N!}\int d^N\sigma\;\prod_{j=1}^N e^{-i\pi k\sigma_j^2}
 \prod_{i<j}^N(2\sinh\pi(\sigma_i-\sigma_j))^2
 \nonumber \\ &&\hskip17mm\cdot
 \bigg(\prod_{j=1}^Ns_{b=1}(i-iq-\sigma_j)\bigg)^{N_f}
 \bigg(\prod_{j=1}^Ns_{b=1}(i-iq+\sigma_j)\bigg)^{\bar N_f}.
\end{eqnarray}

\section{An Application}\label{sec:application}

As a simple application of our result, we consider here an ${\cal N}=4$
SQED with two electron hypermultiplets, called $T[SU(2)]$, which is
long known to be self-mirror \cite{IS,KS}.
In terms of ${\cal N}=2$ supermultiplets, the theory consists of one
abelian vectormultiplet $V$, one neutral chiral multiplet $\phi$ and
four chiral multiplets $(q_1,q_2,\tilde q^1,\tilde q^2)$
with charges $(+1,+1,-1,-1)$. The fields $q_i,\tilde q^i$ have the
R-charge $1/2$ whereas $\phi$ has R-charge $1$. In addition to the
kinetic Lagrangians, a superpotential $W=\sqrt2\tilde q^i\phi q_i$
needs to be introduced to get ${\cal N}=4$ supersymmetry.

The theory has an $SU(2)$ flavor symmetry which rotates $q_i$ and $\tilde q^i$
as doublets, and the matter fields $(q_1,q_2,\tilde q^1,\tilde q^2)$
have charges $(+1,-1,-1,+1)$ under its $U(1)$ subgroup.
One can turn on the mass for the charged chiral matters via gauging
this $U(1)$ flavor symmetry, so that the mass parameter $\mu$ appears
as the expectation value of a background vectormultiplet scalar.
Under the mirror symmetry, the mass parameter
$\mu$ is mapped to the FI parameter $\zeta$ and vice versa.
In a recent work \cite{HLP}, further mass deformation of this
model has been considered by gauging the $U(1)$ symmetry under which
$q_i,\tilde q^i$ all have charge $-1$ and $\phi$ has charge $2$.
Since this symmetry is identified with the difference of two $U(1)$'s
in the $SU(2)\times SU(2)$ R-symmetry, the corresponding mass parameter
$m$ is sign-flipped under the mirror symmetry \cite{Tong}.

The partition function of mass-deformed theory on $S^3$ is thus given by
an integral over the scalar $\sigma$ in the vectormultiplet $V$
(here we set $\ell=1$ for simplicity),
\begin{eqnarray}
 Z(m,\zeta,\mu) &=& \int d\sigma
 e^{4\pi i\zeta\sigma}
 s_{b=1}(-m)\cdot
 s_{b=1}(\tfrac i2-\sigma-\mu+\tfrac m2)
 s_{b=1}(\tfrac i2-\sigma+\mu+\tfrac m2)
 \nonumber \\ && \hskip37mm\cdot
 s_{b=1}(\tfrac i2+\sigma+\mu+\tfrac m2)
 s_{b=1}(\tfrac i2+\sigma-\mu+\tfrac m2).
\end{eqnarray}
By using the following formula given in the appendix of \cite{BT},
\begin{eqnarray}
\lefteqn{
 \int dx e^{-2\pi ipx}
 s_b(x+\tfrac m2+\tfrac{iQ}4)s_b(-x+\tfrac m2+\tfrac{iQ}4)
} \nonumber \\
 &=&
 s_b(m)
 s_b(p-\tfrac m2+\tfrac{iQ}4)s_b(-p-\tfrac m2+\tfrac{iQ}4),
\end{eqnarray}
one can easily show that the partition function satisfies
$Z(m,\zeta,\mu)=Z(-m,\mu,\zeta)$, namely it transforms as expected
under mirror symmetry.

The mass-deformed $T[SU(2)]$ theory is known to describe the S-duality
domain wall of 4D ${\cal N}=2^\ast$ SYM theory\cite{DGG}. As has been
observed in \cite{HLP}, the partition function $Z(m,\zeta,\mu)$
coincides with the S-duality transformation coefficient of Virasoro
torus one-point conformal blocks.

\section*{Note Added}

When our paper was ready for submission to the arXiv, there appeared
a paper by D.L. Jafferis \cite{Jafferis} which has significant overlap
with ours.

\section*{Acknowledgments}
The authors thank Jaemo Park for useful discussions.

\end{document}